\documentclass[%
 reprint,
 amsmath,amssymb,
 aps,
 prl,
]{revtex4-2}

\usepackage{graphicx}
\usepackage{dcolumn}
\usepackage{bm}
\usepackage{color,soul}
\usepackage{enumitem}
\setlist{nosep}

\begin{document}

\preprint{APS/123-QED}

\title{Dynamics of real-time forecasting failure and recovery due to data gaps}

\author{Sicheng Wu}
\author{Ruo-Qian Wang}%
 \email{rq.wang@rutgers.edu}
\affiliation{%
 Department of Civil and Environmental Engineering, Rutgers, The State University of New Jersey, Piscataway, NJ 08854\\
}%

\date{\today}

\begin{abstract}
Real-time forecasting is important to the society. It uses continuous data streams to update forecasts for sustained accuracy. But the data source is vulnerable to attacks or accidents and the dynamics of forecasting failure and recovery due to data gaps is poorly understood. As the first systematic study, a Lorenz model-based forecasting system was disrupted with data gaps of various lengths and timing. The restart time of data assimilation is found to be the most important factor. The forecasting accuracy is found not returning to the original even long after the data assimilation recovery.
\end{abstract}

\maketitle


\emph{Introduction - }Recent decades have witnessed a transition of forecasting systems (e.g. for weather and flood) from the traditional model-centric system to the cyber-physical system that integrates numerical models and monitoring sensor networks: the data from sensor networks is used for model calibration, validation, and data assimilation \citep{beven2018environmental}. As this trend continues,  monitoring networks are becoming increasingly important in key applications such as extreme weather preparation and warning, disaster responses, water supply and irrigation planning, energy, industry, recreational, and ecosystem water uses that are critical to national, state, tribal, and local economic well-being \citep{mason1995stream}.

Such sensor-based monitoring-forecasting systems are being challenged by numerous environmental and social issues, causing data stream disruptions and threatening the system accuracy and reliability. External forcing, such as extreme weather, wildfire, vandalism, and telecommunication issues could compromise the  availability and quality of data to harm the reliability of forecasts in important applications \cite{usgs2018}. Unscheduled maintenance due to cyber-security attacks \cite{gruss2014,dodaro2018}, lack of maintenance due to budget cuts \cite{lundquist2018} and shutdowns caused by large-scale health emergencies also continuously threaten the availability and reliability of the system.

The impact of the monitoring data stream disruptions on the accuracy of forecasting is poorly understood. The forecasting method incorporating real-time data is called data assimilation, which continuously integrates new measurements from monitoring networks to update the state of a deterministic forecasting model. The dominating method of data assimilation is Kalman Filter (KF), which was developed in the 1960s for optimal control of systems governed by linear equations, and its variants \citep{vrugt2005improved}. However, KF-based data assimilation schemes were designed without security in mind \cite{wagner2004resilient}. This loophole makes the monitoring-forecasting system vulnerable to accidents and targeted attacks. The wireless communication community developed a few strategies to enhance the resilience of KF-based systems \cite{wagner2004resilient,nashimoto2018}. To the authors' knowledge, all the past studies focused on the accuracy of data transferring and controllability of robotic systems, and no systematic study has been performed to understand the fundamental physics behind the nonlinear system. This study is designed to fill the knowledge gap. Specifically, we aim at answering the following questions:

$\bullet$ How will the forecast error grow when the real-time data stream fails? 

$\bullet$ Will the forecast error return to the level before the data gap after the real-time data stream recovers? 

$\bullet$ Which factors determine the processes of the forecasting break down and recovery?

This Letter is targeted to use a classical theoretical model to study the dynamics of data stream failures and recovery. 

\emph{Dynamics model and data assimilation - } This study focuses on the classical chaos model developed by Lorenz \cite{lorenz1963deterministic}, which represents the simplified atmospheric convection rolls that are sensitive to the initial conditions. The Lorenz model is a highly nonlinear model and has the typical chaotic behavior. It is worth noting that real forecasting systems are designed to be stable and using the Lorenz model to represent the real system is an exaggeration of the operational system behaviors. Nevertheless, we still select the Lorenz model in this study because of the convenience to trigger observable changes.

The governing equation of the Lorenz model can be written as:
\begin{equation}
\begin{split}
    & \frac{dx_1}{dt} = \sigma(x_2 - x_1) \\
    & \frac{dx_2}{dt} = x_1(\rho - x_3) - x_2 \\
    & \frac{dx_3}{dt} = x_1x_2 - \beta x_3 \\
\end{split}
\label{eq:lorenz}
\end{equation}
\\
in which $\rho$, $\beta$ and $\sigma$ are the model parameters consistent with the original configuration, i.e. $\rho=28$, $\beta=8/3$, $\sigma=10$. This model is annotated as $\textbf{x}(t+\Delta t)=f(\textbf{x}(t),\rho,\beta,\sigma)$.

An Ensemble Kalman Filter (EnKF) is used in this study to perform data assimilation. EnKF is a variant of the original KF, in which an ensemble of state members is used to estimate the covariance \cite{katzfuss2016}. Specifically, the first step is to advance an ensemble of state vectors using the mechanics model by inserting the data of the ensemble ($\Hat{\textbf{x}}_{t-1}^i$) into the Lorenz equations (Eq. \ref{eq:lorenz}) to calculate the next time step $\Tilde{\textbf{x}}_{t}^i$. This time advancing continues until the assimilation time step $t = t_a$ is reached, at which moment the state of the model is update following
\begin{equation}
\label{eq:assim}
\Hat{\textbf{x}}_{t_a}^i = (\textbf{I}_n - \Hat{\textbf{K}}_{t_a}\textbf{H}_{t_a})\Tilde{\textbf{x}}_{t_a}^i + \Hat{\textbf{K}}_{t_a}\textbf{y}_{t_a}^i
\end{equation}
where $\textbf{H}_{t_a}$ is the observation matrix, $\Hat{\textbf{K}}_{t_a}$ is the Kalman gain matrix estimated using the ensemble. Next, a resampling step is performed to regenerate an ensemble of new state vectors based on the estimated covariance to continue the next cycle of data assimilation. Data Assimilation Research Testbed (DART) \cite{anderson2009} is used to implement the data assimilation. The details are listed in Table \ref{tab:dart_setup}.

\begin{table}[ht]
    \centering
    \begin{tabular}{|c|c|}
    \hline\hline
        Numerical Method & Runge-Kutta $2^{nd}$ order  \\ \hline
        Forecasting time step ($\Delta t$) & 0.001 \\ \hline
        Assimilation time step ($\Delta t_a$) & 0.006 \\ \hline
        Total assimilation steps & 10000 \\ \hline
        Ensemble size & 80 \\
    \hline\hline
    \end{tabular}
    \caption{Setup of the data assimilations with DART.}
    \label{tab:dart_setup}
\end{table}

A perfect run, $\textbf{X}_0$, a reference run with data assimilation of continuous data, $\textbf{X}_R$, and assimilated runs with data gaps, $\textbf{X}_G$, are performed, where $\textbf{X}=\{\textbf{x}(0), \textbf{x}(\Delta t), \textbf{x}(2\Delta t), ...\} $ is the time series of the modeling result. The perfect simulation $\textbf{X}_0$ is used as the ``ground truth'' data, which is considered the accurate physics to measure the forecasting error. Because no model is perfect to capture all the details of the physics to make precise predictions, we built a perturbed Lorenz model by increasing $\beta$ of the perfect model by a small value of $10^{-14}$ to represent the systematic error of the forecasting model, i.e. $f_P=f(\textbf{x},\rho,\beta+10^{-14},\sigma)$. Synthetic observational data, \textbf{$\textbf{X}_{obs}$}, is generated by adding a Gaussian random sampling process with a variance of 8.0 to \textbf{$\textbf{x}_{truth}$}, i.e. $\textbf{X}_{obs}=\textbf{X}_0+\epsilon_0$, where $\epsilon_0\sim N(0,8^2)$. The first data assimilation run, $\textbf{X}_R$, was obtained by running $f_P$ with the assimilation of $\textbf{X}_{obs}$ for a total length of 60 time units ($\Delta t=0.006$) without any data gap to serve as the reference. The second run, $\textbf{X}_G$, was obtained with a similar procedure but a data gap was created by removing the observation data from $T_{start}$ with different lengths, $T_L$. Specifically, the procedure can be summarized in three phases: 
\begin{enumerate}
    \item ``\textbf{Normal}'' phase -- the forecasting model, $f_P$, was run to generate a time series of forecast by assimilating the synthetic observation \textbf{$\textbf{X}_{obs}$};
    \item ``\textbf{Gap}'' phase -- $f_P$ was continued without assimilating the synthetic observation starting from $T_{start}$;
    \item ``\textbf{Recovery}'' phase -- $f_P$ was continued at $T_{end} = T_{start} + T_L$ but with the assimilation of the synthetic observation \textbf{$\textbf{X}_{obs}$} again, which was designed to recover the observation data stream. 
\end{enumerate}

In this study, a series of experiments was conducted to generate $\textbf{X}_G$ with $T_{start}$ spreading from 3.3 to 9.3 (non-dimensional time) and $T_{end}$ spreading from $T_{start} + 0.6$ to $10.8$.

The prediction errors are defined as the ensemble average of the square errors between the predictions, $f_G(\textbf{x})$, and the ground truth, $f_0(\textbf{x})$, for the reference ($\epsilon_R$) and the run with gap ($\epsilon_G$) respectively: \cite{atencia2017analogs}:
\begin{equation}
     \epsilon(t) = \frac{1}{N}\sum_{i=1}^{N}[\textbf{x}^i(t) - \textbf{x}_{0}^i(t)]^2,
\end{equation}
where $\textbf{x}$ and $\textbf{x}_0$ are respectively the prediction and ground truth data at $t$, and the index $i$ stands for each of the $N$ ensembles ($N$=80).

The error departure is defined as the difference between $\epsilon_G$ and $\epsilon_R$:
\begin{equation}
\label{eq:dept_gap}
    \Delta_G = \epsilon_G - \epsilon_R
\end{equation}
\\
which measures how much the data gap impacts the data assimilation result.

\begin{figure}[ht]
    \centering
    \includegraphics[width=0.95\linewidth]{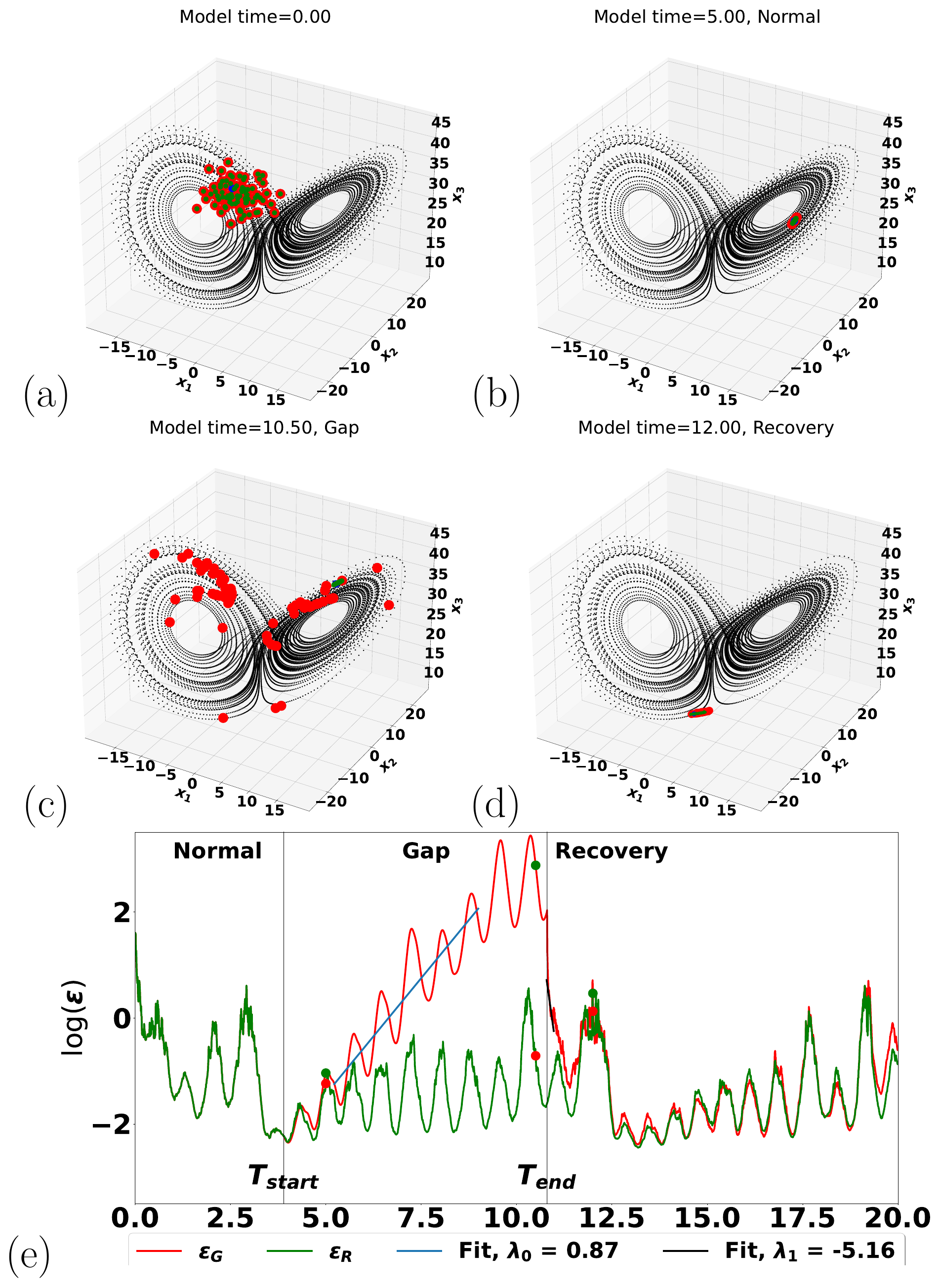}
    \caption{Sample results for data assimilation with and without data gap for $T_{start}$ = 3.9 and $T_{end}$ = 10.8 for (a) T = 0; (b) T = 5.0 before the data gap; (c) T = 10.5 within the data gap; (d)  T = 12.0 shortly after the recovery of the data stream; (e) $\epsilon_G$ and $\epsilon_R$. The red dots are the the results with the data gap, green dots are the results without data gap, and the green dots in (a) -- (d) are the results with.}
    \label{fig:err_gap_sample}
\end{figure}

\emph{Results - } The typical behavior of the data assimilated Lorenz Model is demonstrated in Figure \ref{fig:err_gap_sample}(a)--(d). The red and green dots represent the ensemble location of the ``gap'' and ``reference'' cases, $\textbf{X}_G$ and $\textbf{X}_R$. The ensemble was generated according to the covariance at the beginning of the simulation (Figure \ref{fig:err_gap_sample}(a)). When the data gap begins, the ensemble was concentrated to the same place according to the small covariance (Figure \ref{fig:err_gap_sample}(b)). In the Gap phase (Figure \ref{fig:err_gap_sample}(c)), the ensemble of the reference (green dots) kept a narrow spread thanks to the data assimilation, but the ensemble of the ``gap'' case widely spread because no observation can be used to ``correct'' the model state. Once the data stream was restored, the spread of the ensemble shrank again and the ``gap'' case quickly approached the reference (Figure \ref{fig:err_gap_sample}(d)). This process is quantified in Figure \ref{fig:err_gap_sample}(e), where $\epsilon_G$ and $\epsilon_R$ are compared over the process. The errors of both cases were exactly the same until the beginning of the data gap. Then, $\epsilon_G$ grew exponentially while $\epsilon_R$ remained as in the ``Normal'' phase. In the recovery phase, $\epsilon_G$ significantly dropped to approach $\epsilon_R$ to match despite a small discrepancy.

A series of numerical experiments were performed with a range of $T_{start}$ and $T_L$. We found that the slope of the exponential growth, which is known as the Lyapunov exponent (denoted ``$\lambda_0$''), is always around 0.87 for $\epsilon_G$ in the ``Gap'' phase using linear fitting. The maximum value of $\epsilon_G$ through all the cases was found at the level of $10^3$.

$\epsilon_G$ quickly decreased towards $\epsilon_R$ after the recovery of the data assimilation. To quantify the process, we use the minimum linear slope of $\epsilon_G$ based on the moving linear fitting starting 5 data points after $T_{end}$ with a step of 3 data points until the forecasting recovery ends at $T_{rec}$ -- the first time the forecasting error returns to the reference error, i.e. when $\Delta_G$ reaches 75\% of the $\Delta_G$ at $T_{end} + 20$. The most negative slope or the minimum Lyapunov exponent is denoted $\lambda_1$ to characterize the recovery rate of the system. 

\begin{figure}[h]
    \centering
    \includegraphics[width=0.95\linewidth]{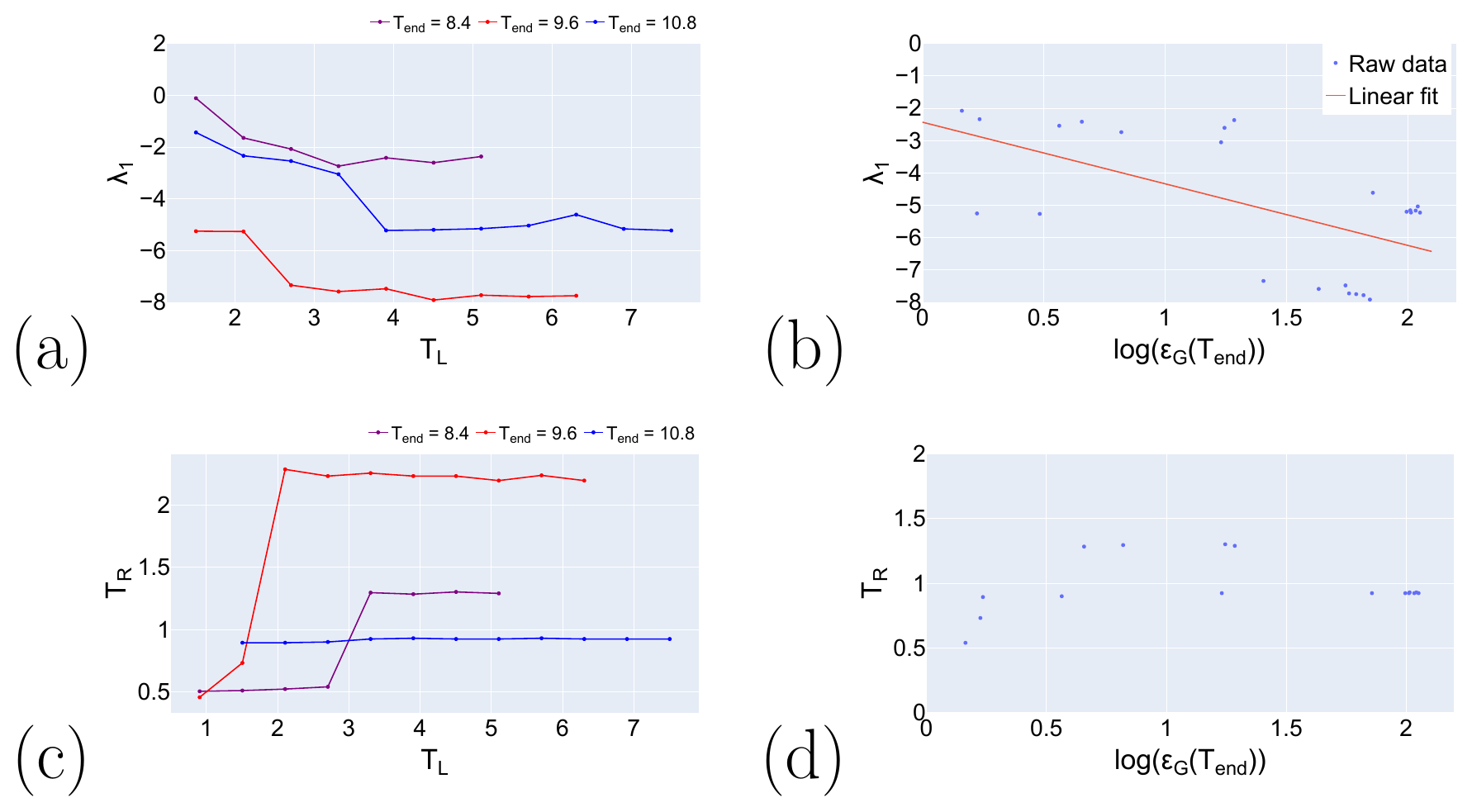}
    \caption{The dependence of the recovery rate $\lambda_1$ on (a) the data missing length, $T_L$, and ((b) the accumulated error at $T_{end}$, and the dependence of the recovery time, $T_R$, on (c) the data missing length, $T_L$, and (d) the accumulated error at $T_{end}$.}
    \label{fig:lambda1}
\end{figure}

$\lambda_1$ is plotted against $T_L$ in Figure \ref{fig:lambda1}(a) to find important factors that impact it. This figure shows that when $T_L$ was relatively small, $\lambda_1$ decreases with $T_L$, while when $T_L$ is large enough $\lambda_1$ weakly depends on $T_L$ but $T_{end}$ plays a more significant role. A further analysis shows that $\lambda_1$ declines exponentially with the forecasting error at the beginning of the data stream recovery, $\epsilon_G(T_{end})$ (Figure \ref{fig:lambda1}(b)). This indicates that the greater the forecasting error, more quickly the forecasting accuracy restores.

While Lyapunov exponents indicate how quickly the forecasting error is corrected, one may be interested in how much time a system needs to recover. So we define the recovery time $T_R$ as the time period between $T_{end}$ and $T_{rec}$:
\begin{equation}
\label{eq:tr}
T_R = T_{rec} - T_{end}
\end{equation}

$T_R$ is found increasing with $T_L$ when $T_L$ is small and being constant with $T_{end}$ when $T_L$ is large (Figure \ref{fig:lambda1}(c)). Regarding the position of the data assimilation restarts, $T_R$ is found increasing with $\epsilon_G(T_{end})$ when $\epsilon_G(T_{end})$ is small, but it kept constant when $\epsilon_G(T_{end})$ is greater than $10^{1.5}$ (Figure \ref{fig:lambda1}(d)). These observation could inform a strategy to recover the data stream for improved forecasting accuracy and more discussion is detailed at the end of this Letter.


\begin{figure}[ht]
    \centering
    \includegraphics[width=0.95\linewidth]{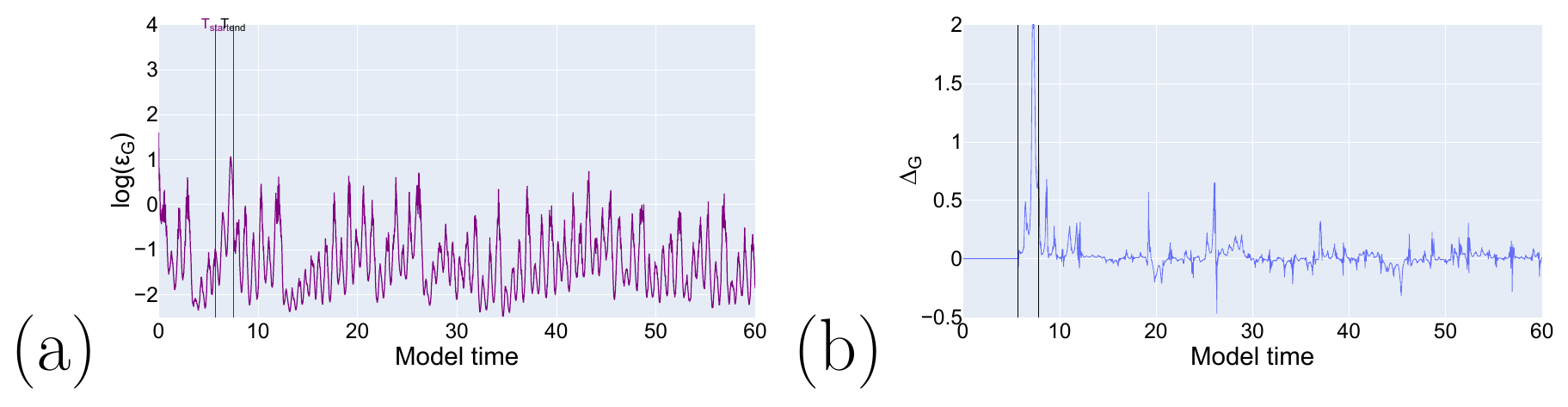}
    \caption{A sample of time series of (a) $\epsilon_G$ and (b) $\Delta_G$ with $T_{start} = 5.7$ and $T_{end} = 7.8$ to show the forecasting errors long after the restart of the data assimilation.}
    \label{fig:dept_gap}
\end{figure}

An interesting observation is made in Figure \ref{fig:err_gap_sample}(e) that despite $\epsilon_G$ quickly returned to the reference once the data assimilation recovered, it never exactly restored to $\epsilon_R$ even long after $T_{end}$. This phenomenon is better reflected in the error departure $\Delta_G$ for a sample case with a short $T_L$ shown in Figure \ref{fig:dept_gap}. It shows in the Recovery phase $\epsilon_G$ in most time exceeded $\epsilon_R$ ($\Delta_G>0$) while it did become lower than $\epsilon_R$ in several instances (i.e. $\Delta_G<0$). Three strong sporadic departure peaked around of 25\% of the maximum $\Delta_G$ during the data gap are observed. The first peak is observed during the recovery of $\epsilon_G$ shortly after $T_{end}$. The two subsequent sporadic departure peaks occurred after $\Delta_G$ returned to near-zero at $T = 12.0$. Some departure peaks of smaller magnitude also occurred after the two strong departure peaks. This demonstrates that the data gap had a permanent disruption to the state of the Lorenz model and a ``hot'' recovery of the data assimilation could not fully restore the accuracy of the original forecast. In average, the mean error of $\Delta_G$ after $T_R$ is 4.94\% $\pm$ 3.50\% based on the standard deviation, with an maximum of 12.7\% across all cases.

\emph{Discussion - } The Lyapunov exponent of around 0.9 in the Gap phase is consistent with previous studies, e.g., \cite{nese1987calculated}. This is expected because without data assimilation the system became a perturbed Lorenz system and the typical bifurcation occurred. This further indicates that a nonlinear, chaotic system will exponentially accumulate the forecasting error if the real-time data stream is lost and the rate of the error accumulation is dependent on the stability of the mechanistic model.

The strange attractors in the phase diagram limited $\epsilon_G$ to the range about $10^{3.2}$. It took about 800 time steps for any case to reach this boundary and the error stopped growing after this saturation (Figure \ref{fig:err_gap_sample}(d)). This observation is consistent with the known characteristics of the system, i.e. the predictability of the Lorenz model is completely lost at around 1000 time steps \citep{atencia2017analogs}.

\begin{figure}[ht]
    \centering
    \includegraphics[width=0.95\linewidth]{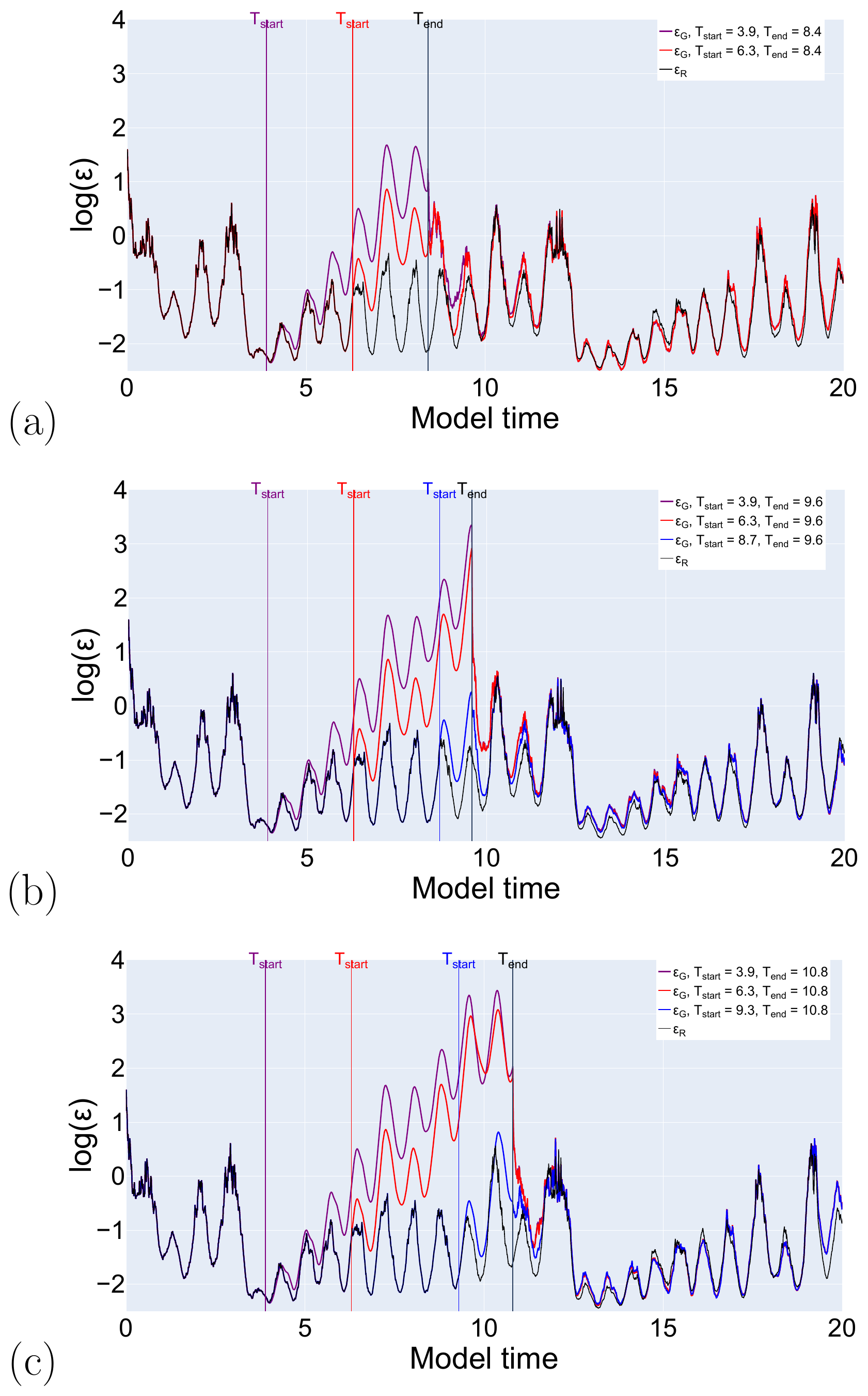}
    \caption{Time series of $\epsilon_G$ for (a) $T_{end} = 8.4$; (b) $T_{end} = 9.6$; (c) $T_{end} = 10.8$ for the same set of $T_{start}$, compared to $\epsilon_R$.}
    \label{fig:err_gap_comparison}
\end{figure}

The saturation point is also shown in the analysis of $\lambda_1$ and $T_R$. At the right end of Figure \ref{fig:lambda1}(b) and (d), the error reached the saturation point so that the recovery processes are similar. While the recovery process before the saturation depends on how much error accumulates when the data assimilation restores. As an analogy, the system takes a ``spring'' style to recover -- the greater the accumulated error, the steeper and longer the recovery process takes. This can be demonstrated in Figure \ref{fig:err_gap_comparison}: with the same $T_{end}$, $\lambda_1$ decreases and $T_R$ increases with the accumulated error until a threshold that the recovery processes converged.

These discoveries could inform forecasting operators to restore a real forecasting system after a data stream failure. First, in a bounded forecasting system, the recovery of the data assimilation will follow the same pattern when the data gap is long enough, i.e. the saturation point is reached. Second, if the data gap is not long enough to reach the error saturation, the operators should fix the system as soon as possible to reduce the full recovery time. Third, because the forecasting error is still present long after the data streams are restored, the restored forecasting system may be different from the perfect forecasting that has continuous data streams. To fully restore the forecasting accuracy, it is necessary to rerun the data assimilation with continuous and consistent data streams.


In summary, this Letter is probably the first endeavor to systematically examine the dynamics of failure and recovery of real-time forecasting systems due to data gaps. Data gaps of various lengths and timing were created in a Lorenz model-based data assimilation system. We discovered that the forecast error grows exponentially with a Lyapunov exponent of around 0.9 because after losing the data assimilation the system becomes a typical chaotic model and experiences a bifurcation. After the observational data recovers, the prediction error declines with a rate faster than the growth of error and the recovery rate and time are positively correlated with the length of the data gap until saturation. 
Forecast error is still observed even long after the recovery of the data assimilation.



\bibliography{apssamp}

\end{document}